\documentstyle[12pt]{article}
\newlength{\dinwidth}
\newlength{\dinmargin}
 \newcommand\beq{\begin{equation}}
 \newcommand\eeq{\end{equation}}
 \newcommand\beqn{\begin{eqnarray}}
 \newcommand\eeqn{\end{eqnarray}}
 \newcommand{\doublespace} {
 \renewcommand{\baselinestretch} {1.6}
 \large\normalsize}

\begin{document}
\vspace*{1cm}
\hspace*{9cm}{\Large MPI H-V29-1996}
\vspace*{3cm}

\begin{center}
{
\renewcommand{\thefootnote}{\fnsymbol{footnote}}

{\Large {\bf Hadronization in Nuclear
  Environment}\footnote{
To appear in the Proceedings of the Workshop on Future Physics at HERA,\\
DESY, September 25, 1995 -- May 31, 1996}}
}
\vspace*{5mm}

 {\large Boris~Kopeliovich$^{ab}$, Jan
  Nemchik$^{cd}$ and Enrico Predazzi$^c$}
\end{center}
$^a$ Max-Planck Institut f\"ur
 Kernphysik, Postfach
103980,  69029 Heidelberg, Germany\\
$^b$ Joint Institute
 for Nuclear Research,
Dubna, 141980
 Moscow Region, Russia\\
$^c$ Universit\`a di  Torino
 and INFN, Sezione di Torino, I-10125, Torino, Italy\\
$^d$Institute of Experimental Physics SAV,
  Solovjevova 47, CS-04353 Kosice, Slovakia\\
\vspace*{1cm}
{{
\doublespace
\begin{quotation}
\noindent
{\bf Abstract:}
We present a space-time description of 
hadronization of highly virtual quarks
originating from
a deep-inelastic electron scattering (DIS).
Important ingredients of our approach are 
the time- and energy--dependence of 
the density of energy loss for gluon radiation,
the Sudakov's suppression of no radiation, and
the effect of 
color transparency, which suppresses final state 
interaction of the produced colorless wave packet.
The model is in a good 
agreement with available data on leading hadron production
off nucleons and nuclei.
The optimal energy range for study of the
hadronization dynamics with nuclear target is 
found to be a few tens of GeV, 
particularly energies available in the experiment HERMES.
\end{quotation}
\newpage

% \section{Introduction}
 
A quark originated from
a hard process,
converts into colorless hadrons owing to
confinement.  Lorentz dilation stretches
considerably the duration
of this process.
While 
hadrons carry a little information about the early
stage of hadronization,  a nuclear target, as a set
of scattering centres, allows us to
look inside the process at
very short times after it starts.  
The quark-gluon system produced in a hard collision, 
interacts while passing through the nucleus.  
This can yield
precious
information about the structure
of this system and the space-time
pattern of hadronization.
 
The modification of the quark fragmentation
function in
nuclear matter was considered for
high-$p_T$ hadron production in
\cite{r2,r9}, for
deep-inelastic lepton scattering in
\cite{r10,kn1,kn2}, and for
hadroproduction of leading particles on
nuclei in \cite{r3,r13}.
The data are usually presented in the form,
% ------------------------------------------
\beq
R_{A/N}=\frac{D_{eff}(z_{h},p_T)}{A\; D(z_{h},p_T)},
\label{1}
\eeq
% ------------------------------------------
where $D(x,p_T)$ and
$D_{eff}(x,p_T)$ are the quark fragmentation function in
vacuum  and in a nucleus, respectively.
 
%\section{Radiative energy loss in vacuum}
 
We treat 
hadronization of a highly virtual quark, 
perturbatively a gluon bremsstrahlung and
the deceleration of the quark as a result 
of radiative energy loss. We assume that
subsequent hadronization of the radiated gluons,
which includes the nonperturbative stage,
does not affect the energy loss of the quark.

The radiation of a
gluon takes the time
% -----------------------------------------------
 \beq
t_r\approx \frac{2\nu}{k_T^2}\alpha(1-\alpha)\ .
 \label{2}
 \eeq
% -----------------------------------------------
This expression follows from 
the form of the energy denominator
corresponding to a fluctuation of a quark of energy $\nu$
into a quark and a gluon, having transverse momenta $k_T$
and relative shares of the initial light-cone momentum
$\alpha$ and $1-\alpha$, respectively. If one calculates radiated
energy taking into account condition (\ref{2}) one
arrives at the density of energy loss per unit of length,
which turns out to be energy and time independent \cite{n}
like in the string model.

In the case of inclusive production of
leading particles at $z_h \rightarrow 1$,
however, energy conservation forbids
the radiation of
gluons with
energy greater than $(1-z_h)\nu$. 
Then, the time dependence of
the radiative energy loss can be written as
% ---------------------------------------- 
 \beq
\Delta E_{rad}(t)=
 \int_{\lambda^2}^{Q^2} dk_T^2
 \int_0^1 d\alpha\ \alpha
\nu
 \frac{dn_g}{d\alpha dk_T^2}
 \Theta(1-z_h-\alpha)\ \Theta(t-t_r)\, ,
\label{5}
 \eeq
% ----------------------------------------
where $dn_g/d\alpha dk_T^2 = \epsilon /\alpha k_T^2$ 
represents
the distribution of the number of gluons.
The factor 
$\epsilon=4\alpha_{S}(k_T^2)/3\pi$.

Although soft hadronization is usually described
in terms of the string model,
we model it by radiation as well, choosing the
bottom limit $\lambda^2$ in (\ref{5}) small.
We fix the QCD running coupling $\alpha_{S}(k_T^2) =
\alpha_{S}(k_0^2)$ at $k_T^2 \leq k_0^2$, in the region
which is supposed to be dominated by nonperturbative
effects. the parameter $k_0 \approx 0.7\
GeV$ is chosen
to reproduce the density of energy loss 
for radiation of soft gluons ($k_T^2 \leq k_0^2$)
corresponding
to the string tension, $dE/dt=
\rangle \approx 1\ GeV/fm$. This value of $k_0$ is consistent with
the transverse size of a string corresponding to 
the gluon-gluon correlation
radius, $R_{c}\sim 1/k_{0}\sim 0.3$\,fm
suggested by QCD lattice results.

After integrating 
eq. (\ref{5}) in the soft radiation 
approximation, we get
% ----------------------------------------- 
\beqn
 & & \Delta E_{rad}(t)=
 {\epsilon\over 2} t (Q^2-\lambda^2)
\Theta(t_1-t) +
 \epsilon\nu (1-z_h)\Theta(t-t_1) + \nonumber\\
 & &
\epsilon\nu (1-z_h) \ln\left ({t\over t_1}\right )
 \Theta(t-t_1)
\Theta(t_2-t_1) +
 \epsilon\nu (1-z_h) \ln\left ({Q^2\over \lambda^2}
 \right
) \Theta(t-t_2)
 \label{6}
 \eeqn
% -----------------------------------------
Here we have set
$t_1=(1-z_h)/x_{Bj}m_N$ and 
$t_2=t_1\ Q^2/\lambda^2$,
where $x_{Bj}$
is the Bjorken variable.

Eq. (\ref{6}) shows that for $t\le t_1$,
the density of energy loss is constant, 
$dE/dt=-\epsilon Q^2/2$, exactly
as in the case with no restriction on the radiated
energy \cite{kn1,kn2}.
At longer time intervals, $t> t_{1}$
more energetic gluons can be
radiated and the restriction $\alpha <
1-z_h$ becomes important.  
As a result, the density of energy loss slows
down to $dE/dt=-\epsilon\nu (1-z_h)/t$
which is a new result compared to what was known
in the string model.
At still longer $t> t_{2}$, no radiation is
permitted, but obviously a color charge 
cannot propagate a long time without
radiating which must be 
suppressed by a Sudakov's type formfactor.
Assuming a Poisson distribution for the number of
emitted gluons we get the
formfactor,
$
F(t)=\exp\left[-\tilde n_g(t)\right]
$, 
where $\tilde n_g(t)$
is the number of non radiated gluons,
% ------------------------------------------------- 
\beq
\tilde n_g(t)=\epsilon
\left[{t\over t_1}-1
 -\ln\left ({t\over t_1}\right )\right ]
\Theta(t-t_1) \, .
\label{77}
\eeq
% -------------------------------------------------

In order to calculate a time interval for 
the leading hadron production
(or, better, a colorless ejectile
which does not loose energy anymore), one needs 
a model of hadronization and
of color neutralization.
In the large $N_c$ limit,
each radiated gluon can be replaced 
by a $q\bar q$ pair,
and the whole system can be treated as a
system of color dipoles.
It is natural to
assume that 
the leading (fastest) hadron
originates from a $q\bar q$ dipole
made of the leading quark and of the
antiquark coming from the last
emitted gluon.
This dipole is to be projected
into the hadron wave function, $\Psi(\beta,l_T)$,
where $\beta$ and $1-\beta$ are the
relative shares of the light-cone momentum carried by the
quarks, and $l_T$ is the relative transverse
momentum of the quarks.
The result of this projection leads to the fragmentation
function of the quark into the hadron, which reads
% --------------------------------------------
\beq
D(z_h) = \int_0^{\infty} dt W(t,\nu,z_h)\ ,
\label{88}
\eeq
% --------------------------------------------
where $W(t,\nu,z_h)$ is a distribution function 
of the leading hadrons over the production time $t$.
% --------------------------------------------
\beqn
 W(t,\nu,z_h)& \propto &
 \int \limits_0^1
\frac{d\alpha}{\alpha}\
 \delta \left
 [\alpha-2\left
(1-\frac{z_h\nu}{E_q(t)}\right )\right ]
 \int \frac{dk_T^2}{k_T^2}\
\delta\left
 [k_T^2-\frac{2\nu}{t}
 \alpha(1-\alpha)\right ]\times
\nonumber\\
 & & \int dl_t^2\ \delta\left [l_T^2-{9\over
 16}k_T^2 \right ]
\int\limits_0^1 d\beta
 \delta\left [\beta-\frac{\alpha}{2-\alpha}
 \right ]
|\Psi_h(\beta,l_T)|^2 \, .
 \label{9}
 \eeqn
% ---------------------------------------------- 
Here the quark energy $E_q(t) = \nu-\Delta E_{rad}(t)$.
We have chosen a hadronic wave function in the
light-cone representation which satisfies the
Regge end-point behaviour, $|\Psi_h(l_T^2,\beta)|^2
\propto \sqrt{\beta} \sqrt{1-\beta} (1
+ l_T^2r_h^2/6)^{-1}$, 
where $r_h$ is the charge radius of the hadron.

Fig.1 shows function $W(t)$
for several values of $z_{h}$ and exhibits 
the approximate $(1-z_h)\nu$-scaling  
of the mean production time,
$t_{pr}=\int dt\ t\ W(t)$,
which depends weakly on $Q^2$ and
vanishes at $z_{h}\approx 1$. 

Our predictions
for the fragmentation function $D(z_h)$
depicted in Fig.~2 nicely agrees 
with the EMC data
\cite{emc}.

}}

% ===================================================================
\begin{figure}[ht]
\includegraphics{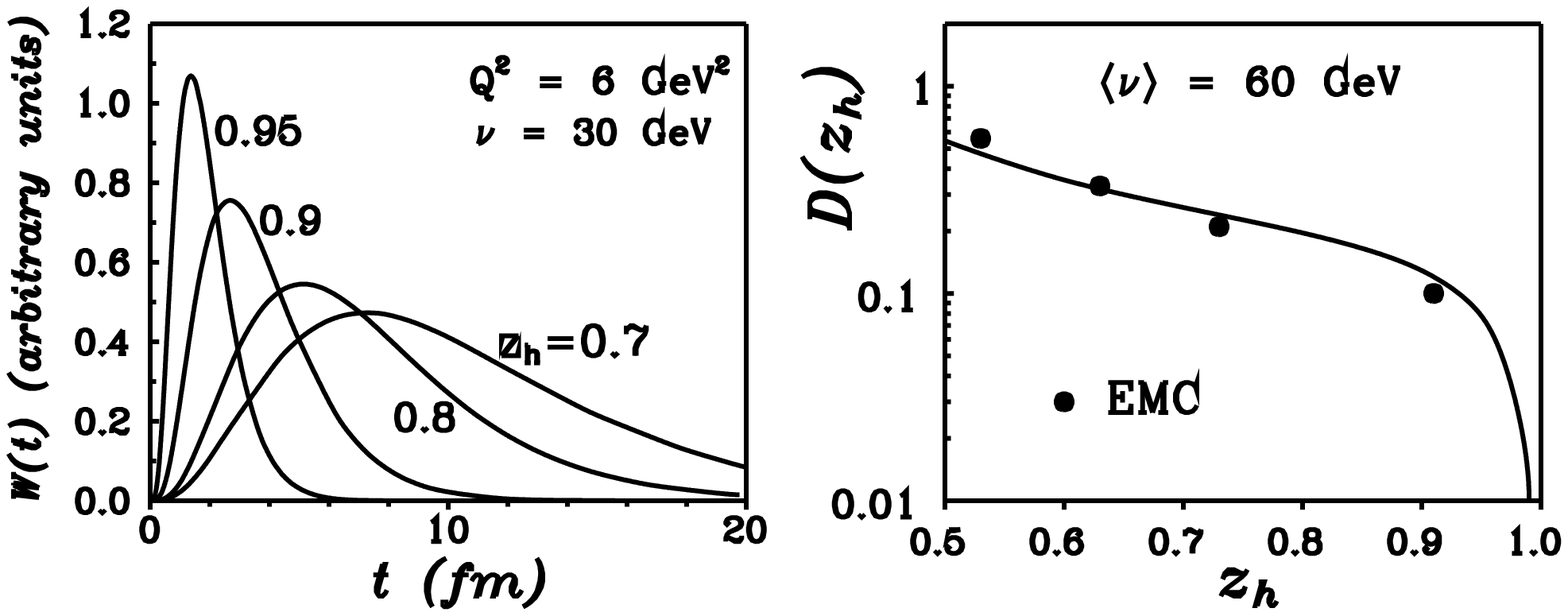}
\begin{center}
\vspace{5.5cm}
\parbox{13cm}
{\caption{\it Distribution of the hadron production 
time at $\nu=30\ GeV$, $Q^2=6\ GeV^2$ 
and $z_h = 0.70 \div 0.95$}
\caption{\it Comparison of our prediction for $D(z_{h})$ with
data [13]}}
\end{center}
\end{figure}
% ===================================================================

\begin{figure}[ht]
\includegraphics{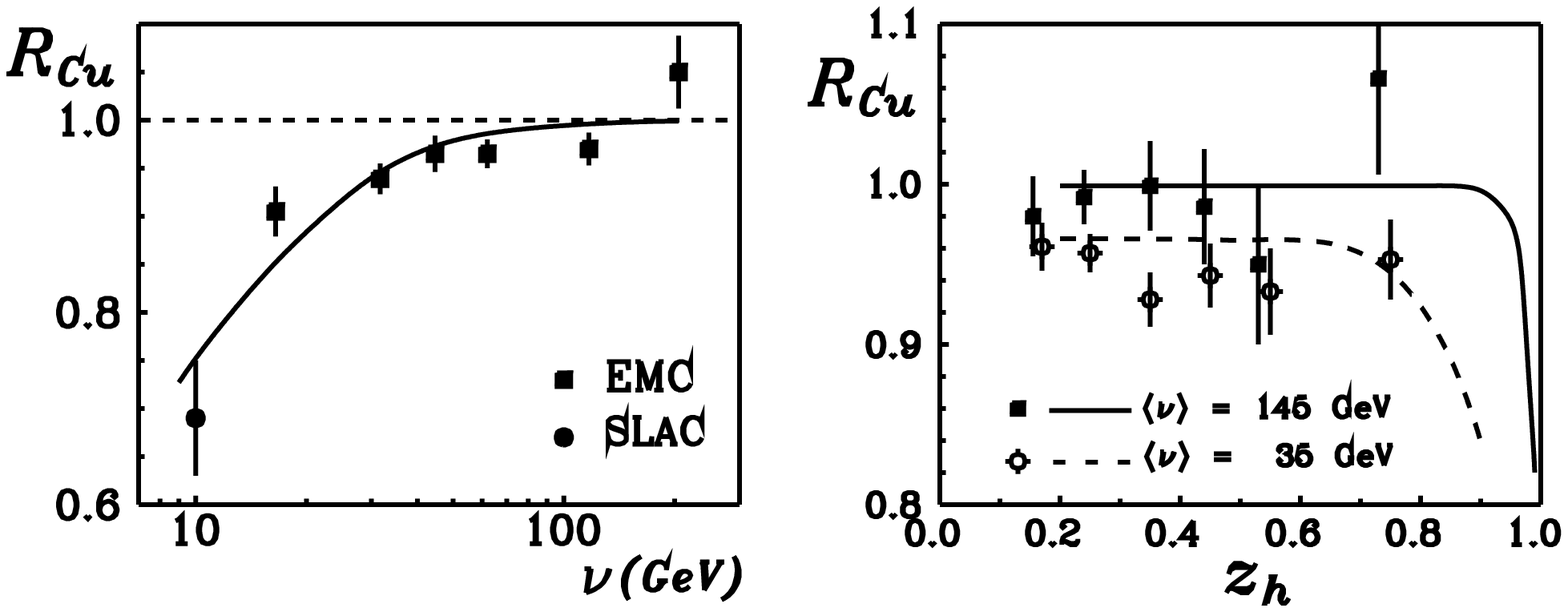}
\begin{center}
\vspace{5.5cm}
\parbox{13cm}
{\caption{\it Comparison of our calculations for the $\nu$-dependence
of nuclear suppression (1) integrated over $z_h$
at $Q^2=6\ GeV^2$ with
data [13,14]} 
\caption{\it $z_h$-dependence of 
nuclear suppression.}}
\end{center}
\end{figure}
% ===================================================================

\begin{figure}[t]
\includegraphics{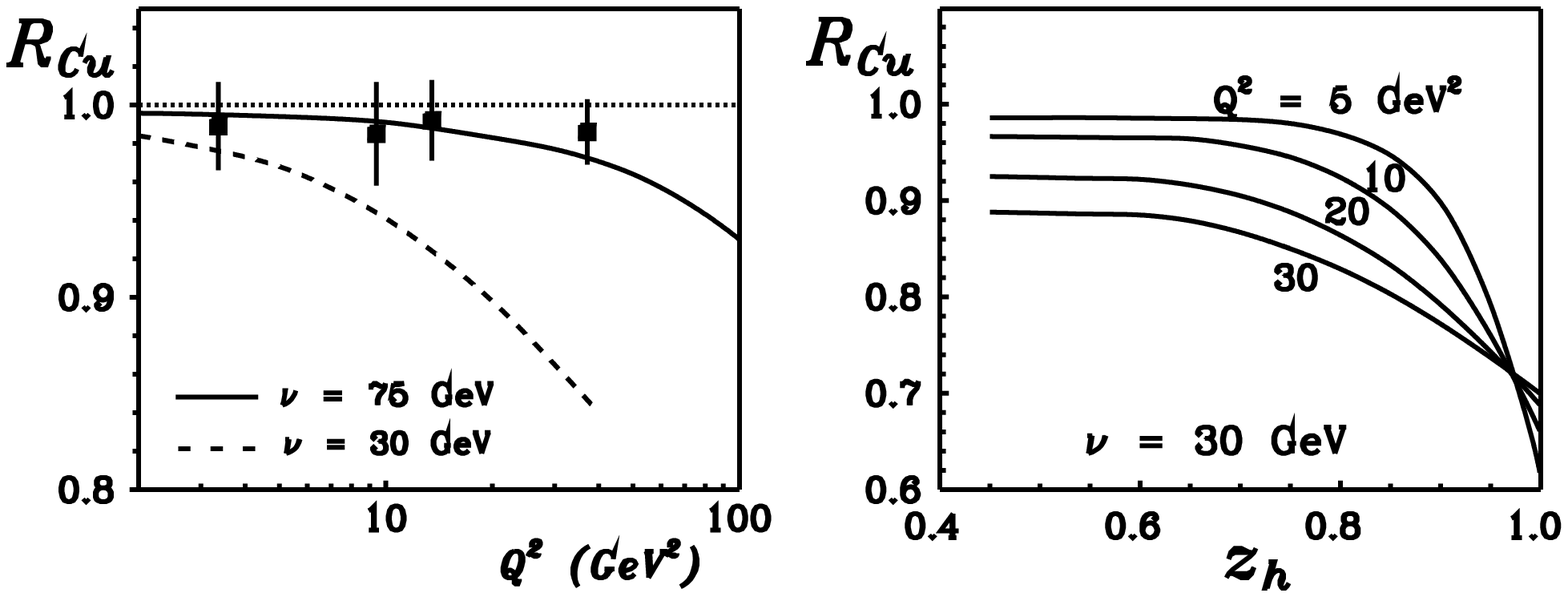}
\begin{center}
\vspace{5.5cm}
\parbox{13cm}
{\caption{\it Predictions for the $Q^{2}$ dependence of 
nuclear transparency at $\nu = 30$ and $75\ GeV$.
Data [14] at $\nu = 75\ GeV$ are integrated over $z_{h}$}
\caption{\it Predictions for the $z_h$ dependence
of nuclear transparency at $\nu = 30\ GeV$
vs $Q^2$}}
\end{center}
\end{figure}
% ===================================================================

\vspace*{1cm}
{{
\doublespace

%\section{Hadronization in nuclear medium}

The
production of the leading colorless wave packet 
with the desired (detected)
momentum 
completes the process of hadronization.
Any subsequent inelastic interaction is
forbidden, otherwise a new hadronization process
begins and the leading
hadron energy falls to lower values.
Such a restriction means
a nuclear suppression of the production rate.

On the other hand, soft
interactions of the leading
quark during the hadronization in
nuclear matter
cannot stop or absorb the leading quark \cite{k90}.
Although rescatterings of the
quark in the nucleus result in
an additional induced soft radiation,
\cite{k90,gw}, at the same time
the quark looses much more energy due to the
hard gluon radiation following the deep-inelastic
scattering just as in vacuum.  
Thus, the induced soft radiation can be treated as
a small correction to the energy loss and can be
neglected, provided $Q^2$ is high enough.
 
The transverse size
of the colorless wave packet
produced in a hard reaction can be small,
therefore the nuclear
suppression is weaker
due to color transparency.
We take into account
the evolution of the wave packet during its
propagation through the nucleus using the path
integral technique developed in \cite{kz91}.
Figs.~3 - 5 show quite a good agreement
of our parameter-free calculations 
with available data \cite{slac,emc}
on the $\nu$-, $z_h$- and $Q^2$-dependence of
nuclear suppression.
Unfortunately, there is still no data 
in the region, $z_h>0.8$.

Note that at high energies 
many of interesting effects
go away or are difficult to observe.
Nuclear suppression integrated over $z_h$ vanishes
(see Fig.~3).
The region of high
$z_h$, where nuclear suppression is expected to be 
enhanced (see Fig.~4) squeezes at high $\nu$ 
towards $z_h=1$, where
the cross section vanishes.
There is almost no $Q^2$-dependence at high $\nu$ and 
moderate values of $Q^2$ (see Fig.~5).

We present in Figs.~5 and 6 our 
predictions for the $Q^2$ and $z_{h}$ 
dependence of nuclear suppression 
for the energy range of the HERMES experiment.
We expect the onset 
of nuclear effects  at moderate values of 
$Q^{2}$ (Fig.~5) as well as 
in the region of $z_{h} > 0.8$ (Fig.~6),
which can be tested by the HERMES experiment.
 
\medskip
 
To conclude, we have developed a phenomenology of
electroproduction of
leading hadrons on nuclei
 which is based on the perturbative QCD.  
Our parameter-free model is in a 
good agreement with
available data.
We stress that the energy range of the HERMES
experiment is especially sensitive to 
the underlying dynamics of hadronization.

}}

\end{document}